\documentclass[preprint]{emulateapj}

\slugcomment{Draft version 24. November 2006}

\shorttitle{High resolution spectroscopy of GRB 030329}
\shortauthors{C. C. Th\"one et al.}

\def\farcs{\hbox{$.\!\!^{\prime\prime}$}} 
\begin{document}

\title{ISM studies of GRB 030329 with high resolution spectroscopy \footnote{Based on ESO proposal 70.D-0087}}

\author{Christina C. Th\"one \altaffilmark{1}, Jochen Greiner, Sandra Savaglio}
\email{cthoene@dark-cosmology.dk}
\affil{Max-Planck-Institut f\"ur Extraterrestrische Physik, Giessenbachstra\ss e, 85748 Garching, Germany}

\author{Emmanu\"el Jehin}
\affil{European Southern Observatory, Alonso de Cordova 3107, Vitacura, Santiago 19, Chile}

\altaffiltext{1}{DARK Cosmology Centre, Niels-Bohr-Institut, Juliane Maries Vej 30, 2100 K\o benhavn \O, Denmark}

\begin{abstract}
We present a series of early UVES/VLT high resolution spectra of the afterglow of GRB
030329 at redshift $z=0.16867\pm0.00001$. In contrast to other spectra from this burst, both emission and absorption lines were detected. None of them showed any temporal evolution.  From the emission lines, we determine the properties of the host galaxy which has a star formation rate (SFR) of 0.198\,M$_{\odot}$ yr$^{-1}$ and a low metallicity of 1/7\,Z$_\odot$. Given the low total stellar host mass M$_\star$=10$^{7.75\pm0.15}$\,M$_\odot$ and an absolute luminosity m$_\mathrm{V}$=--16.37, we derive specific SFRs (SSFR) of log SFR/M = --8.5 yr$^{-1}$ and SFR/L = 14.1 M$_{\odot}$ yr$^{-1}$ L$_*^{-1}$. This fits well into the picture of GRB hosts as being low mass, low metallicity, actively star forming galaxies.\\
The Mg\,{\sc ii} and Mg\,{\sc i} absorption lines from the host show multiple narrow (Doppler width b=5-10 km/s) components spanning a range of v$\sim$260 km/s, mainly blueshifted compared to the redshift from the emission lines. These components are likely probing outflowing material of the host galaxy, which could arise from former galactic superwinds, driven by supernovae from star forming regions. Similar features have been observed in QSO spectra. The outflowing material is mainly neutral with high column densities of log N(Mg\,{\sc ii})=${14.0\pm0.1}$ cm$^{-2}$ and log N(Mg\,{\sc i})=${12.3\pm0.1}$ cm$^{-2}$.
\end{abstract}


\section{Introduction}
Gamma-Ray-Bursts (GRBs) have proven to be one of the lighthouses in the distant universe since the determination of their cosmological origin in  1997 \citep{Paradijs97, Metzger97}. Their afterglow spectra, ranging from X-ray to radio, follow a power law decay and do not seem to show intrinsic lines from the burst itself though there are some claims of X-ray line detections from the prompt emission \cite[e.g.,][and references therein]{Butler05}. This makes them well suited as a powerful light source to study the burst environment in the host galaxy and the medium in the line of sight by using absorption lines. In addition, emission lines from the host might be seen on top of the afterglow spectrum and provide important information about the type of galaxy in which GRBs occur. As these galaxies are usually very faint and difficult to detect, GRBs offer a unique possibility to find and study these galaxies in detail.\\
One of the few other possibilities to study these faint, distant galaxies is through QSO observations \citep{Ellison, Petitjean00}. Their spectra often show intervening absorption systems from foreground damped Lyman alpha systems (DLAs) or metal absorption lines such as Mg\,{\sc ii} \citep{Churchill03} associated with galaxies or halos of galaxies. Mg\,{\sc ii} absorbers are the best studied sample to date as Mg\,{\sc ii} is the reddest of the strong metal absorption lines in the interstellar medium (ISM), but also C\,{\sc iv}, which probes a different region of the ISM has been studied since many years. Recently, highly excited lines like O\,{\sc vi} have also been discovered in such absorption systems \citep{Ellison00, Schaye00}. With high resolution spectra, the stronger absorption lines of these QSO-absorbers usually have been found to consist of several components spanning velocities up to 300\,km/s \citep{Prochter06}. These different components are assumed to originate in different clouds in the absorbing system and are therefore able to give a picture of the ISM structure in these galaxies.\\
For an increasing fraction of GRBs, spectra of the afterglow can be obtained, especially since the {\it Swift} satellite \citep{Gehrels2004} started operating in December 2004. Depending on the redshift and the wavelength range of the instrument, these spectra may contain emission lines from HII regions in the host galaxy or absorption lines from material inside the host galaxy and/or the intergalactic medium (IGM). Emission line analysis from the host galaxy can be easily performed with low resolution spectroscopy from the afterglow containing emission lines from the galaxy, or with later observations targeting the galaxy. These analyses reveal that host galaxies of long GRBs are usually young, irregular, low mass, low metallicity, star-forming galaxies \citep{Christensen}. To study the ISM structure in the host galaxy through absorption lines, however, higher resolution is needed which is only available for few bursts so far \citep[and references therein]{Savaglio06}, most of them detected recently. Their absorption lines usually show several components with a wide range of velocities whose nature is still speculative. Due to the wavelength coverage of the spectra and the position of the absorption and emission lines, there are only very few bursts that contain both. Examples are GRB\,990712 \citep{Vreeswijk04} and GRB\,970508 \citep{Reichart98} which had Mg\,{\sc ii} absorption in addition to the host emission lines. GRB\,020813 \citep{Barth03} showed weak OII emission in addition to the broad absorption lines containing many velocity components and GRB\,020405 \citep{Masetti03} had a weak Ca\,{\sc ii} absorption line. Only for one burst, GRB\,060218 \citep{Wiersema}, high resolution spectra containing both kind of lines could be obtained which show different components in the host absorption and one of the emission lines.
\\ 
\\
The long GRB 030329 was detected by the HETE-2 satellite at 11:38:41 UT \citep{VanderspekGCN} and lasted for $T_{90}=22.9$\,s at high energies. In terms of fluence, it was among the top 1\% of all detected GRBs. A bright optical afterglow was detected 1.5 hr after the burst with an R-band magnitude of 12.5 \citep{Price} allowing to get high resolution spectra. The optical lightcurve showed a slow decay with a power law index of $\alpha=-1.2\pm0.1$ and several phases of rebrightenings 1.3, 2.4, 3.1 and 4.9 days \citep{Matheson03, Lipkin04} after the onset of the burst. From the emission lines of the host in the UVES spectra presented here, a redshift of z=0.1685 was determined \citep{JochenGCN}.\\
Due to its brightness and low redshift, a series of low resolution spectra could be taken over several weeks, which showed for the first time the connection between long-duration GRBs and supernovae (SNe) \citep{Hjorth03, Matheson03, Stanek03}. Also, for the first time, it was possible to measure the time evolution of the afterglow polarization \citep{JochenNature}. Emission line analysis of the host galaxy was obtained from an extensive low resolution dataset taken with the FORS spectrograph \citep{Hjorth03, Sollerman05} and spectra from \cite{Matheson03} who found a a SFR of 0.5M$_{\odot}$/yr and a moderate metallicity of 12+log(O/H)=8.5, but they assumed the upper branch of the metallicity solution.  None of them however was able to detect any absorption line as the wavelength range of their spectra missed the Mg II lines in the UV.\\
In this paper, we present high resolution UVES/VLT spectra from the first and the fourth night after the burst. We detect and analyze both emission and absorption lines as shown in \S 3. The emission lines are used to derive the redshift, extinction, SFR and metallicity of the host galaxy in \S 4.  \S 5 shows the splitting of the host absorption lines into 7 components and gives a possible interpretation of these components.\\
For the calculations in the paper, we used a $\Omega_{\Lambda}$ = 0.7, $\Omega_{m}$ = 0.3 cosmology, a value of H=70 km/s for the Hubble constant and c=299,792 km/s.

\section{Data sample}

\begin{figure*}
\includegraphics[scale=0.9, angle=0]{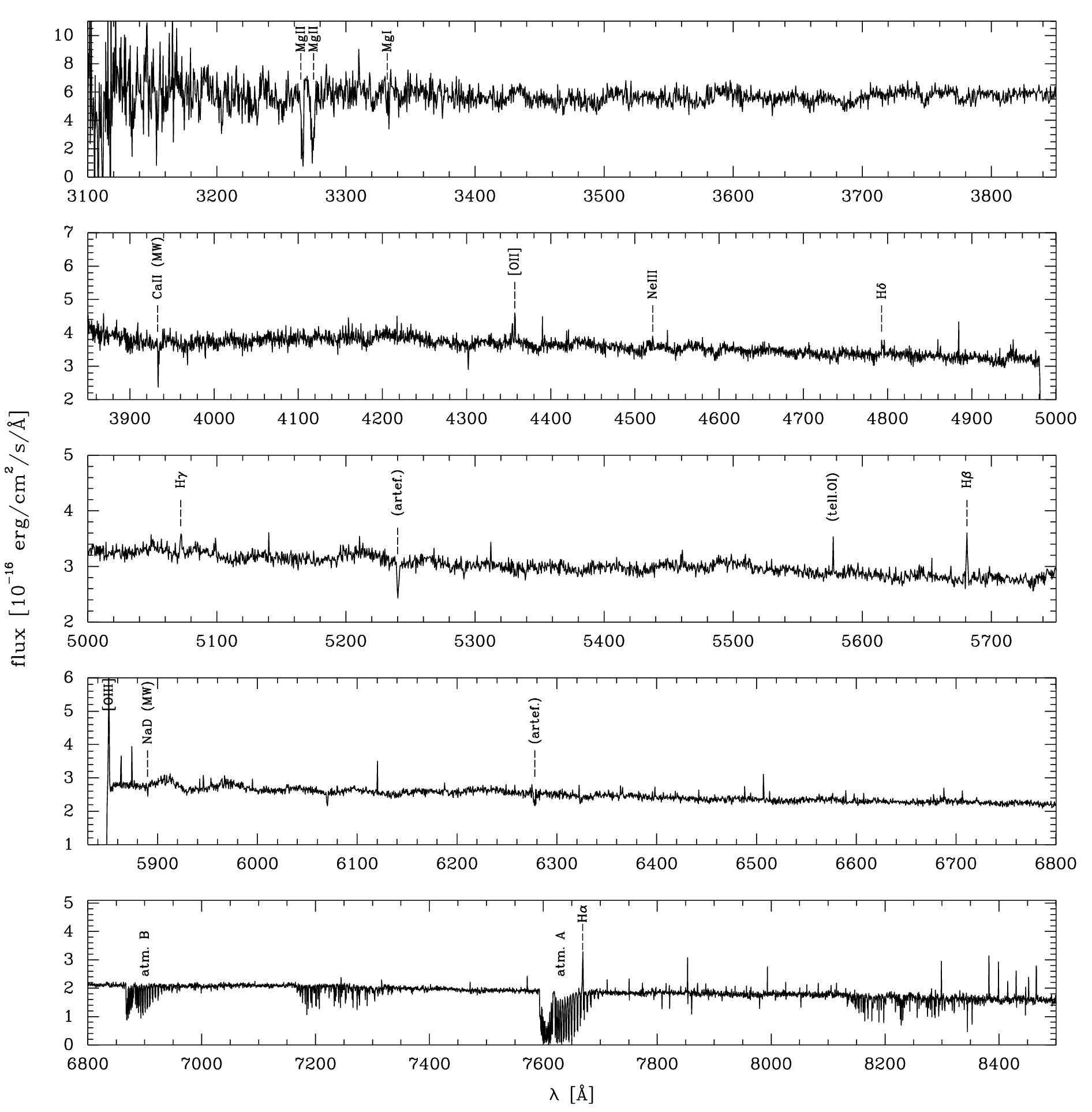}
\caption{Total UVES spectrum of the fourth night after the burst where the afterglow continuum has faded away enough to result in a good S/N of the host emission lines. For better
visualization of the spectral lines in this graph, the spectra have been smoothed and rebinned to 0.4 \AA{}. Ca\,{\sc ii} and NaD lines are from absorption in the Milky Way (MW), all other absorption and emission lines are from the host galaxy at z=0.16867. The emission features not marked in panel 3 and 4 are due to imperfect atmospheric emission line removal as their FWHM is much smaller than from the lines of the host galaxy and they are only visible in the spectra of this epoch. The two features at 5238 and 6069 \AA are due to bad pixels on the CCD.
Note that the scale of the y-axis changes for the different panels.
\label{totalspectrum}}
\end{figure*}

High-resolution spectra of the afterglow of GRB 030329 have been collected with the UV-Visual Echelle Spectrograph \citep[UVES,][]{Dekker00}, of the ESO Very Large Telescope (VLT), on the nights of March 30 (UT) and April 2, 2003 under good seeing (0\farcs9-0\farcs7) conditions. Four 30 minutes exposures and two 1 hour exposures were secured on the first and fourth night after the burst, respectively, using two different beam splitters (dichroic \#1 and \#2). Two UVES standard settings were used, the so-called DIC1 (R346 nm, B580 nm), and the DIC2 (R437 nm, B860 nm). This combination allowed us to get on the first observing date two spectra covering the full optical range (303-1060 nm), except for small gaps near 575 and 855 nm, and one similar spectrum on the second night. The use of a 0\farcs8 slit width provides a resolving power of $\lambda/\Delta\lambda$ $\sim$ 55,000 or 5.5 km/s (FWHM). The slit was oriented along the parallactic angle in order to minimize losses due to the atmospheric dispersion.\\
The raw data were reduced using the UVES pipeline (v1.4) implemented in the ESO-MIDAS software package
\citep{Ballester00}. This pipeline reduction includes flat-fielding, bias and sky subtraction and a relative wavelength calibration which has an accuracy of $\pm$ 0.5 km/s. The optimum extraction method has been used, which assumes a Gaussian profile for the cross-dispersion flux distribution and is optimized for low SNR spectra. The individual reduced spectra were also corrected for the airmass, the atmospheric absorption bands could however not be removed because of saturation. No correction for vacuum wavelengths or heliocentric velocities (-16.23 and -17.36\,km/s for the two nights respectively) was done for the final spectra, this was taken into account later for the redshift determination (see Sec. \ref{results}).\\
We also did an absolute flux calibration using the master response curves determined by the ESO Garching Quality Control\footnote{http://www.eso.org/observing/dfo/quality/UVES/qc/\\std\_qc1.html\#response} \citep[see also][]{Hanuschik03}. Comparison with photometric data of the afterglow from the literature \citep{Matheson03} showed that the continuum of the flux calibrated spectrum was almost a factor 2 too low. This can hardly be explained with bad seeing It seems, however, to be a common problem in UVES flux calibration. Comparison with FORS spectra taken at the same time suggest that the UVES flux calibration is about a factor of 1.6 too low \citep{Wiersema}. We therefore recalibrated the spectrum using photometric data from \cite{Matheson03}, obtained at the same time as our spectroscopic observations.\\
The log of the observations are noted in Table \ref{logobs}. As an example, the entire reduced spectrum of the third epoch is shown in Fig.
\ref{totalspectrum}.\\

\begin{deluxetable*}{lllllllll}
\tablewidth{0pt} 
\tablecaption{Log of the observations}
\tablehead{\colhead{epoch} &\colhead{date} &\colhead{UT} &\colhead{d after GRB} &\colhead{exp. [min]} &\colhead{R mag}& \colhead{seeing ['']} &\colhead{airmass}&\colhead{wavelength range [nm]}}
\startdata
1&30/3/2003&03:22:44&0.6556&30&15.27&0.73&1.45&303-388 + 476-684\\
&&03:55:43&0.6785&30&15.31&1.03&1.49&373-499 + 660-1060\\
2&30/3/2003&04:29:01&0.7016&30&15.38&0.97&1.59&303-388 + 476-684\\
&&05:01:53&0.7244&30&15.46&0.75&1.74&373-499 + 660-1060\\
3&02/4/2003&00:46:54&3.5474&60&17.24&0.49&1.72&383-499 + 660-1060\\
&&01:51:51&3.5925&60&17.25&0.72&1.49&303-388 + 476-684\\
\enddata
\label{logobs}
\tablecomments{Dates given are the start of the observations}
\end{deluxetable*}

\section{Line analysis}

\subsection{Emission lines}
\begin{figure}
\includegraphics[scale=0.88, angle=0]{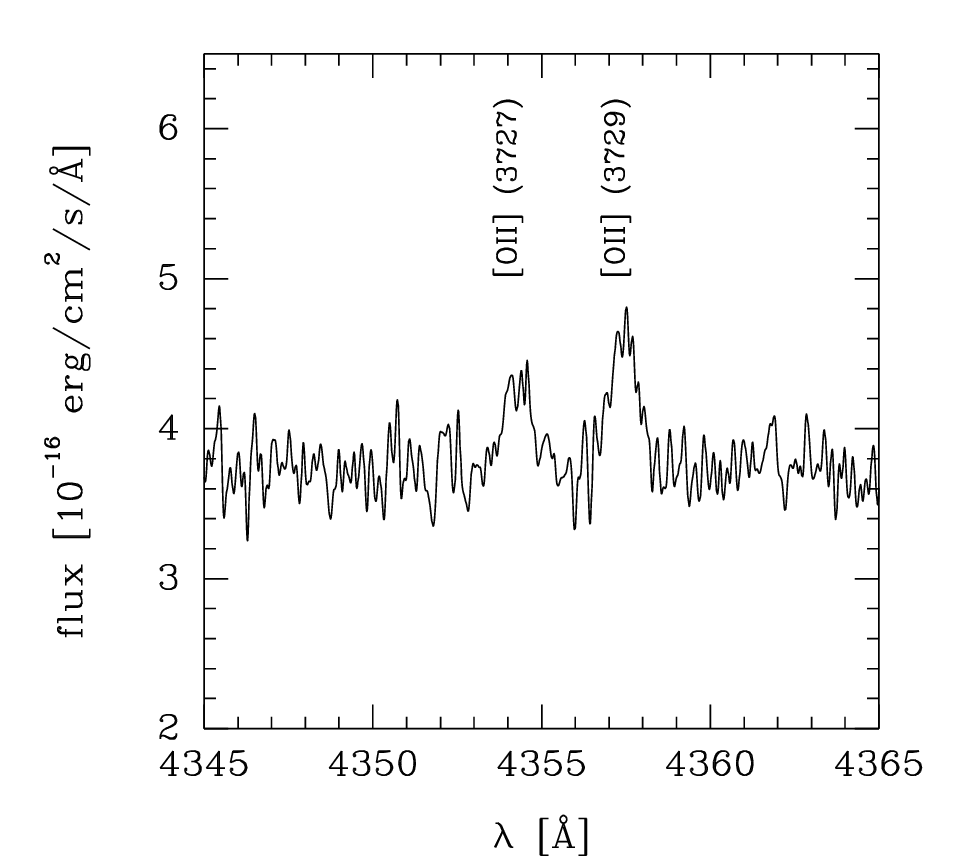}
\includegraphics[scale=0.88]{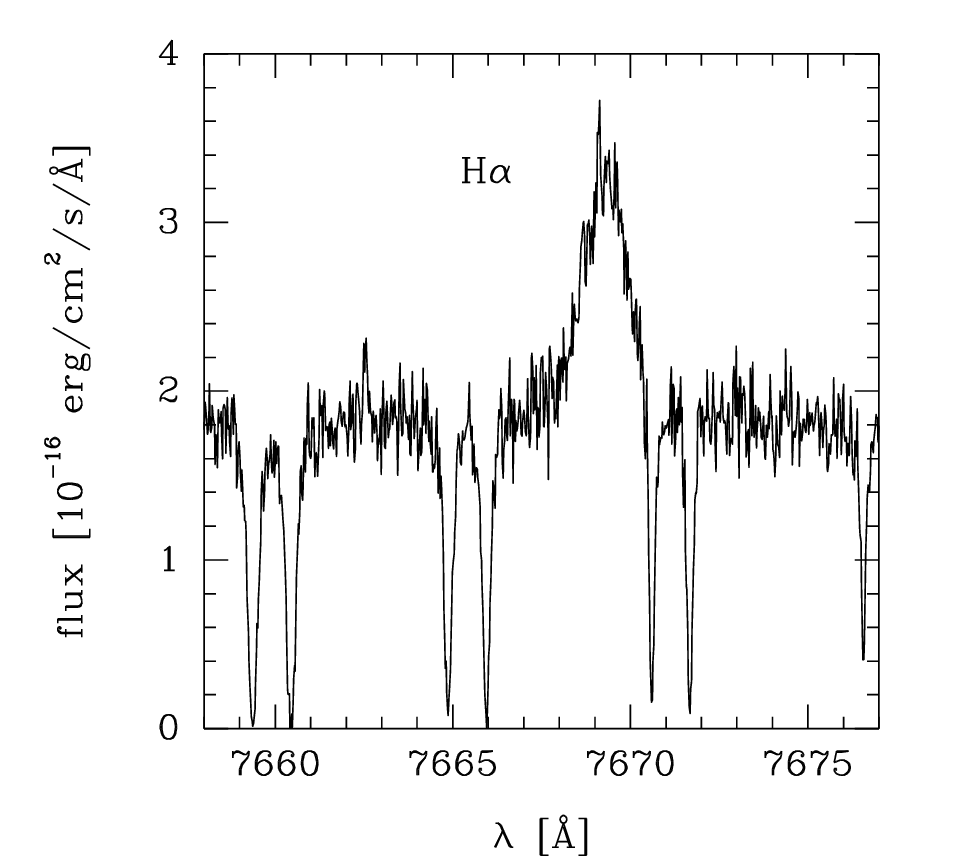}
\caption{Upper figure: Resolved OII lines in the spectra from the 3rd epoch. For better visualization, the spectrum has been smoothed with a box car width of 10 pixels. Lower figure: Position of the H$\alpha$ line in between two atmospheric absorption lines.
\label{OII}}
\end{figure}

We detected a range of emission lines from the GRB host galaxy. The Balmer series down to H$\delta$, the forbidden lines [OII]\,$\lambda\lambda$3727,3729 and [OIII]\,$\lambda$5007 as well as [Ne III]\,$\lambda$3869 could be identified. H$\alpha$\, was also detected though it lies in the middle of the atmospheric A absorption band. Fortunately, it fell between two of the resolved atmospheric absorption lines in the atmospheric A band and is therefore very little affected by absorption (see Fig. \ref{OII}). H$\delta$\, and  H$\gamma$\, could only be detected clearly in the data from the third epoch. Of the two [OIII] lines at 4963 \AA{} and 5007 \AA{}, the 4963 \AA{} line unfortunately falls in the gap of the detector in the red arm which consists of a mosaic of two chips. [N II]\,$\lambda$6586 was not detected, but a reliable 2$\sigma$ upper limit could be derived because of the high resolution, despite the fact that this region in the spectrum was also affected by atmospheric absorption bands.\\
The [OII] doublet was well resolved in the spectra of the fourth night (see Fig. \ref{OII}) which was only possible in three other bursts, GRB\,990506 and GRB\,000418 \cite{Bloom03} as well as GRB\,060218 \citep{Wiersema}. This is especially interesting because the [OII]\,$\lambda\lambda$3727/3729 ratio allows the determination of the electron density n$_e$ \citep{Osterbrock89}. In our spectrum, the ratio of the two lines is 0.53$\pm$0.10 for the third epoch when the S/N was high enough to resolve the 3727 \AA-line, because the afterglow continuum had faded away. This ratio is consistent with the lower value of 0.66 for n$_e\rightarrow 0$. Similar ratios were found for the hosts of GRB\,990506, GRB\,000418 and GRB\,060218 with values of 0.57$\pm$0.14, 0.75$\pm$0.11 and 0.62$\pm$0.05 respectively.
\noindent
To measure the emission line fluxes, we fitted the lines with a Gaussian and compared it with measuring the line flux directly which gave similar results. The properties of the emission lines in the individual spectra as well as the mean value of the flux from first night's spectra are listed in Table \ref{emission}. The flux of the emission lines is generally higher in the two spectra from the first night (epoch 1 and 2) which could be due to different seeing and it also varies between the two exposures. As the light curve of the first night showed large variations and even small color changes \citep{Matheson03, Lipkin04}, this might also affect the spectrum. Therefore, we take the emission line fluxes from the 3rd epoch for further calculations as the continuum might be less influenced by intrinsic changes of the afterglow spectrum.\\
Even after recalibrating the spectra with photometric data, our emission line fluxes are a factor of 1.3 (second epoch) to 1.4 (first epoch) lower than the values found by \cite{Gorosabel},\cite{Sollerman05} in late FORS spectra from the host galaxy. As they used a 1\farcs0 slit and a fixed orientation angle covering most of the galaxy, these data might not be entirely comparable to our spectra, though the line ratios are very similar to those found from later spectra.\\

\begin{deluxetable*}{llllllll} 
\tablewidth{0pt} 
\tablecaption{Emission line characteristics}
\tablehead{\colhead{Line} & \colhead{epoch} &
\colhead{$\lambda_{rest}$} & \colhead{$\lambda_{obs}$} &
\colhead{FWHM}  & \colhead{Flux (measured)}& \colhead{Flux (A$_V$ corr.)}& \colhead{Mean(1,2) A$_V$ corr.}\\
\colhead{} & \colhead{} & \colhead{[\AA]} & \colhead{[\AA]} & \colhead {[\AA] and [km/s]} & \colhead {[10$^{-17}$erg/cm$^2$/s/\AA]} &\colhead{[10$^{-17}$erg/cm$^2$/s/\AA]}&\colhead{[10$^{-17}$erg/cm$^2$/s/\AA]}}

\startdata
OII & 1& 3726 	& (4354)  & (0.9)					& $<$17.3$\pm$0.5	&	$<$32.1$\pm$0.9&\\
 & 2& 		& (4354)  & (0.9)					& $<$20.6$\pm$0.5	&	$<$38.2$\pm$0.9& $<$ 35.2\\
 & 3 & 		& 4354.29 & 0.91$\pm$0.10\ \ 64$\pm$7	& 4.81$\pm$0.5 	&	8.92$\pm$0.9	&\\[2mm]

OII & 1& 3728 	& 4357.67 & 0.62$\pm$0.10\ \ 37$\pm$7 	& 11.5$\pm$0.5 	&	21.4$\pm$1.0	&\\
 & 2& 		& 4357.37 & 0.69$\pm$0.10\ \ 40$\pm$7 	& 17.9$\pm$0.5 	&	33.3$\pm$1.0	& 27.3\\
 & 3 & 		& 4357.43 & 0.96$\pm$0.10\ \ 56$\pm$7	& 9.04$\pm$0.5   	&	16.7$\pm$1.0	&\\[2mm]

NeIII& 1& 3869 & 4520.80 & 0.74$\pm$0.10\ \ 42$\pm$6 	& 6.6$\pm$2.0 	 	&	12.1$\pm$3.7	&\\
 & 2& 		& 4521.39 & 0.50$\pm$0.10\ \ 28$\pm$6 	& 8.5$\pm$2.0   	&	15.6$\pm$3.7	& 13.85\\
 & 3 & 		& 4521.27 & 0.91$\pm$0.10\ \ 51$\pm$6	& 3.4$\pm$2.0 	 	&	6.24$\pm$0.9	&\\[2mm]
 
H$\delta$&1&4101& (4793)&(0.9)	 				& $<$16.3$\pm$1.0  &	$<$29.2$\pm$1.8&\\
 & 2& 		& (4793)  	 & (0.9)	 				& $<$17.3$\pm$1.0	&	$<$31.0$\pm$1.8& $<$ 30.1\\
 & 3 & 		& 4793.07 & 0.83$\pm$0.10\ \ 45$\pm$6	& 3.35$\pm$1.5 	&	6.00$\pm$1.8	&\\[2mm]

H$\gamma$&1&4340& (5072) &(0.9)	 			& $<$8.39$\pm$1.0  &	$<$14.6$\pm$1.8&\\
 		   & 2& 	  & (5072)  &(0.9)			 	& $<$11.0$\pm$1.0  & 	$<$19.2$\pm$1.8& $<$16.9\\
 		   & 3 &	  & 5072.21& 0.81$\pm$0.10\ \ 41$\pm$6	& 3.2$\pm$1.5 	 	& 	5.58$\pm$1.8	&\\[2mm]

H$\beta$&1&4861& 5681.17 & 0.90$\pm$0.10\ \ 41$\pm$5 & 10.0$\pm$1.0   &	16.3$\pm$3.2	&\\
 & 2& 		& 5680.86 & 0.95$\pm$0.10\ \ 43$\pm$5 	& 8.4$\pm$1.0		&	13.8$\pm$3.2	& 15.1\\
 & 3 & 		& 5680.99 & 1.00$\pm$0.10\ \ 45$\pm$5	& 7.8$\pm$0.5 	 	&	12.8$\pm$1.7	&\\[2mm]

OIII	&1&5007 & 5850.90 & 0.77$\pm$0.10\ \ 33$\pm$5 	& 39.4$\pm$0.5   	&	63.4$\pm$0.8	&\\
 & 2& 		& 5851.01 & 0.81$\pm$0.10\ \ 36$\pm$5 	& 38.0$\pm$0.5  	&	61.1$\pm$0.8	& 62.3\\
 & 3 & 		& 5851.03 & 0.76$\pm$0.10\ \ 33$\pm$5	& 31.8$\pm$0.5  	&	51.2$\pm$0.8	&\\[2mm]

H$\alpha$&1&6563& 7669.24 & 1.33$\pm$0.10\ \ 45$\pm$4 & 23.5$\pm$1.0 	&	32.9$\pm$1.3	&\\
 & 2& 		& 7669.22 & 1.21$\pm$0.10\ \ 40$\pm$4 	& 26.9$\pm$1.0 	 &	37.6$\pm$1.3	& 35.3\\
 & 3 & 		& 7669.29 & 1.40$\pm$0.10\ \ 47$\pm$4	& 21.3$\pm$0.5 	 &	29.8$\pm$0.8	&\\[2mm]

 N II & 1& 6586 & (7695)  & (1.20) 			& $<$8.77$\pm$1.0		 	& 	 $<$12.2$\pm$1.3&\\
 & 2 & 		& (7695)  & (1.20) 			& $<$9.63$\pm$1.0 			& 	 $<$13.4$\pm$1.3& $<$12.8\\
 & 3 & 	        & (7695)  & (1.20) 			& $<$2.67$\pm$0.5 			& 	 $<$3.71$\pm$0.8&\\
\enddata
\label{emission}
\tablecomments{Values for the emission lines for the three different datasets with the numbers indicating the spectra taken in the first and the fourth night after the burst as described in Table \ref{logobs}. The measured wavelenghts are in air and not corrected for heliocentric velocity, FWHM is restframe corrected. Fluxes are given as the measure value and the value corrected for Galactic and host extinction as described in Sec. \ref{ext}}
\end{deluxetable*}

\subsection{Absorption lines}
Compared to other bursts from which spectra could be obtained, GRB\,030329 has a very low redshift. Therefore, most of the metal absorption lines are still far in the UV, outside the detection range of UVES.\\
However, we detect the Mg\,{\sc ii}\,$\lambda\lambda$2797, 2803 doublet clearly resolved as well as Mg\,{\sc i}\,$\lambda$2852, which all reveal to have several well resolved velocity components. Voigt profiles were fitted to these absorption features using the FIT/LYMAN package \citep{Fontana} in MIDAS and the column densities determined for each component (see Sec. \ref{abskin})\\
Interstellar NaD\,$\lambda$5890 and Ca\,{\sc ii}\,$\lambda$3933 from our Galaxy were also detected and resolved in three different clouds, which allows an independent determination of the Galactic extinction. These two elements do usually not occur in the same regions in the galaxy and their wavelength and properties can therefore be different. We find that the first two clouds have the same shift within the accuracy, the third one however has a different position for NaD and Ca\,{\sc ii}. The FWHM and the ratio of the EWs NaD/Ca\,{\sc ii} are also different within each cloud which favors different states of the ISM for the origin of the two absorption lines \citep{Welty}.\\
We could, however, not detect any NaD\,$\lambda$5890, Ca\,{\sc ii}\,$\lambda$3933 or Ti\,{\sc ii}\,$\lambda$3242 absorption from the host galaxy, whereas the ratio of Ca\,{\sc ii} to Mg\,{\sc i}, which have similar ionization potentials, found in GRB 020813 \citep{Savaglio04} was around 1. Ratios found in the MW can vary between 0.01 and 1 depending on the properties of the ISM \citep{Welty99}\\
The properties of the absorption lines are listed in Table \ref{abscharac}.

\begin{deluxetable}{llllll}
\tablewidth{0pt}
\tablecaption{Absorption line characteristics }
\tablehead{ \colhead{Line} & \colhead{epoch} & \colhead{$\lambda_{rest}$}   &
\colhead{$\lambda_\mathrm{obs}$} & \colhead{FWHM}  & \colhead{EW}\\
\colhead{} &\colhead{} &\colhead{[\AA]}&\colhead{[\AA]}&\colhead{[\AA]}&\colhead{[\AA]}}
\startdata
Mg II & 1 	& 2796	& 3266.1	& 1.7$\pm$0.1		& 1.3$\pm$0.1\\
      	& 2 	&  		& 3266.2 	& 1.8$\pm$0.2		& 1.3$\pm$0.2\\ 
      	& 3 	&   		& 3266.7 	& 1.9$\pm$0.2		& 1.1$\pm$0.2\\[1mm]
Mg II & 1 	& 2803 	& 3274.5 	& 1.9$\pm$0.1 		& 1.3$\pm$0.2\\
      	& 2 	&   		& 3274.7 	& 1.4$\pm$0.2 		& 0.9$\pm$0.2\\
      	& 3 	&   		& 3274.8 	& 2.8$\pm$0.2 		& 1.63$\pm$0.30\\[1mm]
Mg I  & 1 	& 2852 	& 3332.5 	& 1.2$\pm$0.2 		& 0.25$\pm$0.18\\
      	& 2	& 		& 3333.0 	& 1.3$\pm$0.2 		& 0.25$\pm$0.18\\
      	& 3 	&   		& 3332.9 	& 1.9$\pm$0.2		& 0.39$\pm$0.25\\[1mm]\hline
	&	&		&		&				&		\\[-1.5mm]
NaD &1	& 5890 	& 5889.58& 0.23$\pm$0.02 	& 0.053$\pm$0.01\\
	&	&		& 5890.27	& 0.24$\pm$0.05 	& 0.028$\pm$0.01\\
	&	&		& 5890.60& 0.16$\pm$0.05 	& 0.006$\pm$0.01\\
	&2	&		& 5889.64& 0.28$\pm$0.05	& 0.050$\pm$0.01\\
	&	&		& 5890.21& 0.25$\pm$0.05 	& 0.027$\pm$0.01\\
	&	&		& 5890.54& 0.17$\pm$0.05 	& 0.009$\pm$0.01\\
	&3	&		& 5889.64& 0.21$\pm$0.05 	& 0.031$\pm$0.01\\
	&	&		& 5890.01& 0.10$\pm$0.05 	& 0.021$\pm$0.01\\
	&	&		& 5890.15& 0.05$\pm$0.05 	& 0.013$\pm$0.01\\[1mm]
Ca II	& 1	& 3933 	& 3933.5	& 0.25$\pm$0.03 	& 0.16$\pm$0.01 \\
	&	&		& 3933.9 	& 0.22$\pm$0.03 	& 0.05$\pm$0.01\\
	&	&		&3934.6 	& 0.30$\pm$0.03	& 0.04$\pm$0.01\\
	&2	&		& 3933.5	& 0.23$\pm$0.05 	& 0.18$\pm$0.01\\
	&	&		& 3933.9 	& 0.17$\pm$0.05	& 0.03$\pm$0.01\\
	&	&		& 3934.5	& 0.28$\pm$0.05 	& 0.06$\pm$0.01\\
	&3	&		& 3933.5 	& 0.27$\pm$0.05 	& 0.23$\pm$0.01\\
	&	&		& 3933.9 	& 0.13$\pm$0.05 	& 0.03$\pm$0.01\\
	&	&		&3934.5 	& 0.30$\pm$0.05 	& 0.09$\pm$0.01\\
\enddata
\label{abscharac}
\tablecomments{The upper part are absorption lines from the host galaxy, the lower part from the MW. The FWHM of the Mg lines are the widths over the rebinned lines. The different wavelengths and EWs of the NaD interstellar lines from the third epoch are due to imperfect sky line removal from telluric NaD emission.}
\end{deluxetable}

\section{Results}\label{results}
\subsection{Redshift}
The redshift of the host galaxy was determined from the [O III]\,$\lambda$5007, H$\alpha$ and H$\beta$ emission lines of the first epoch \citep{JochenGCN} as $z=0.1685$. We did a refined analysis using the emission lines [O II]\, $\lambda\lambda$3726, 3729, [O III]\,$\lambda$5007, H$\alpha$ and H$\beta$, corrected for heliocentric velocities of $-$16.23 and $-$17.36 km/s for night 1 and 4 respectively. From this, we derive a value of $z=0.16867\pm0.00001$, consistent with the value found by \cite{Matheson03} and \cite{Sollerman05}. Note, however that this is the mean redshift of the GRB host galaxy, not of the GRB itself.

\subsection{Broadening of the emission lines}
Low resolution spectra are usually not able to distinguish between instrumental resolution and intrinsic broadening of the host emission lines due to galaxy rotation and turbulence of the medium.\\
From our high resolution spectra (FWHM=5.5 km/s), we find that all emission lines in our spectra are clearly broadened compared to the resolution of the instrument. The FWHM of the emission lines lies between 35 and 55 km/s, restframe corrected. This fits well to the fact that the host of GRB\,030329 is a dwarf galaxy, which have typical radial velocities up to 100 km/s. We might however not see the whole rotational velocity spread if the galaxy is not seen edge on.

\subsection{Galactic and host extinction}\label{ext}
For the determination of the Galactic extinction, usually sky catalogs such as \cite{Schlegel} or \cite{Dickey90} are used. The strength of interstellar lines like NaD\,$\lambda$5890 however allows a different and more accurate determination of the extinction using e.g. the model of \cite{Munari}. In our spectra, Galactic NaD as well as Ca\,{\sc ii} show a triple absorption peak, indicating three clouds in our Galaxy in the line of sight. The NaD-line gives an extinction of E$_\mathrm{B-V}$ = 0.013 $\pm$ 0.004,  0.007 $\pm$ 0.002 and 0.002 $\pm$ 0.001 for the three clouds respectively which results in a total extinction of E$_\mathrm{B-V}=0.022\pm0.005$ or A$_{\rm{V}}$=0.068 as A$_{\rm{V}}$=R$_{\rm{V}}$E$_\mathrm{B-V}$ with R$_{\rm{V}}$=3.1 for a MW extinction law \citep{CCM}. This is comparable to the extinction determined from the Schlegel sky catalog of E$_\mathrm{B-V}=0.025$.\\
The extinction in the host galaxy can in principle be derived from the Balmer line decrement, the constant flux ratio of the Balmer lines derived from line transition properties, with unextinguished values of H$\alpha$/H$\beta$=2.76 and H$\gamma$/H$\beta$=0.474 for a temperature of 10,000 K and case B recombination \citep{Osterbrock89}. From our spectrum we obtaine H$\alpha$/H$\beta$=2.73$\pm$0.2 which is consistent with no extinction.  The ratio H$\gamma$/H$\beta$=0.41$\pm$0.15 would give an extinction of E(B-V)=0.74$\pm$0.7 which is unfortunately quite inaccurate due to the relatively large error in the emission line fluxes. In addition, the Balmer lines can be affected by absorption from an underlying stellar population depending on the age of the galaxy. This affects H$\beta$ and H$\gamma$ much more than H$\alpha$, which means that the derived extinction should be seen as an upper limit. However, in the UVES spectra, we could not identify any underlying stellar absorption, whereas \cite{Gorosabel} derive a relatively high correction, comparable to the value of the fluxes for H$\gamma$ and H$\delta$ from their stellar population models. We consider it therefore to be more appropriate to use an extinction value derived from the broad band afterglow as it was done in \cite{Kann}. They derive a value of A$_{\rm{V}}=0.39\pm0.15$, fitting an SMC extinction law, which differs however very little from the fit with a MW extinction law.\\
To correct the emission line fluxes, we then use A$_{\rm{V}}=0.39$ for the host extinction and A$_{\rm{V}}=0.068$ for the Galactic extinction applying a MW extinction law (see Table\ref{emission}).

\subsection{SFR}
A direct indicator of the current SFR in a galaxy is the flux from nebular emission lines such as H$\alpha$ that had been ionized by young, massive stars. Using the H$\alpha$ flux from the 3rd epoch, which is corrected for extinction but not for a possible underlying stellar absorption, we derive a SFR of the host galaxy according to \cite{Kennicutt} of SFR [M$_{\odot}$/yr] = $7.9\cdot10^{-42} 4 \pi f_\mathrm{H\alpha} d_L^2$ =  0.198$\pm$0.006 M$_{\odot}$/yr which is consistent with the value in \cite{Hjorth03} of SFR$_\mathrm{H\alpha}$=0.17 M$_{\odot}$/yr from the FORS spectra. Another indicator for the SFR are forbidden lines like [OII] with the drawback of being only indirectly coupled to the ionizing flux of young stars but very sensitive to the ionization state of the ISM. For [OII], the corresponding equation SFR [M$_{\odot}$/yr] = $1.4\cdot10^{-41} 4 \pi f_\mathrm{[OII]} d_L^2$ gives a value of 0.302$\pm$0.006 M$_{\odot}$/yr which is higher than the value derived from H$\alpha$. As the SFR from [OII] seems to be less reliable, we adopt here the SFR from H$\alpha$. \\
A SFR of 0.198 M$_{\odot}$/yr is not very high compared to other galaxies at this redshift, but one has to consider the low mass of the host galaxy as already noted in \cite{Sollerman05}. Savaglio, Glazebrook and Le Borgne (2006, in prep.) derive a low total stellar mass for the host of M$_\star$ = 10$^{7.75\pm0.15}$M$_\odot$. This gives a specific SFR (SSFR) per unit mass of log SFR/M = $-$8.5 yr$^{-1}$. Furthermore, the host galaxy of GRB 030329 is underluminous with an absolute magnitude of m$_\mathrm{V}$=$-$16.37 \citep{Fruchter06}. This can be used to scale the SFR with the luminosity fraction compared to a standard galaxy with m$_\mathrm{V(abs)}$=$-$21 to derive the SFR/L \citep{Christensen}. For our galaxy, we get a value for the SFR/L of 14.1 M$_{\odot}$ yr$^{-1}$ L$_*^{-1}$ which is high compared to the average of other host galaxies of 9.7 M$_{\odot}$ yr$^{-1}$ L$_*^{-1}$ as found by \cite{Christensen}.\\

\subsection{Metallicity of the host}
There are a number of secondary, empirical metallicity calibrators derived from the abundances of nebular emission lines. The most frequently used, because easiest to get, is the R$_{23}={{{f_\mathrm{[OII]\,\lambda3727}}+f_\mathrm{[OIII]\\lambda\lambda4959, 5007}}\over{f_\mathrm{H\beta}}}$ parameter \citep{Kobulnicky}. [O III]\,$\lambda$4959 was not detected due to the gap in the spectrum, there is, however a fixed ratio between the two [O III] lines, f$_\mathrm{[OIII]\,\lambda5007}$=3$\cdot$f$_\mathrm{[OIII]\,\lambda4959}$. \\ 
The R$_{23}$ parameter gives two possible solutions, in our case either a metallicity of 12+log(O/H) = 7.8 or 8.7. This degeneracy can be broken using the [N II]\,$\lambda$6586 to H$\alpha$ ratio \citep{Lilly}. As noted in \cite{Sollerman05}, no [N II] was detected in the spectra of GRB\,030329, except from the FORS spectrum of May 1st \citep{Hjorth03}, which seems to be spurious. Due to the high resolution of the UVES spectrum however, we were able to put a strong 2$\sigma$ upper limit of [N II] (see Table \ref{emission}) and therefore to derive a ratio of [N II]/H$\alpha$ from the strongest upper limit at the 3rd epoch of $< 0.12\pm0.02$. This favors a lower branch solution according to \cite{Lilly}, which gives a low metallicity value of 12+log(O/H) = 7.8 or 1/7 Z$_{\odot}$, assuming 12+log(O/H) = 8.66 for solar metallicity \citep{Asplund04}. The low mass of the host galaxy supports a low value for the metallicity.\\
The R$_{23}$ parameter however has turned out to be affected by systematic errors for high and low metallicities \citep{Bresolin04}. \cite{Wiersema} showed that the metallicity of the GRB 060218 host, derived from the oxygen abundances, is 12+log(O/H)=7.54, compared to 8.0 from the R$_{23}$ parameter. For a direct calibration from the abundances however, the electron temperature and density have to be known which rely on the detection of [OIII]\,$\lambda$4363 and the detection or low error flux measurements of the [OII]\,$\lambda\lambda$3727,3729 or [SII]\,$\lambda\lambda$6716,6731 doublets, which were not detected for the GRB\,030329 host or only with large errors in the case of [OII].

\section{Absorption line kinematics}\label{abskin}
\subsection{Fitting of the absorption systems}
\begin{figure*}
\centering\includegraphics[scale=1.4]{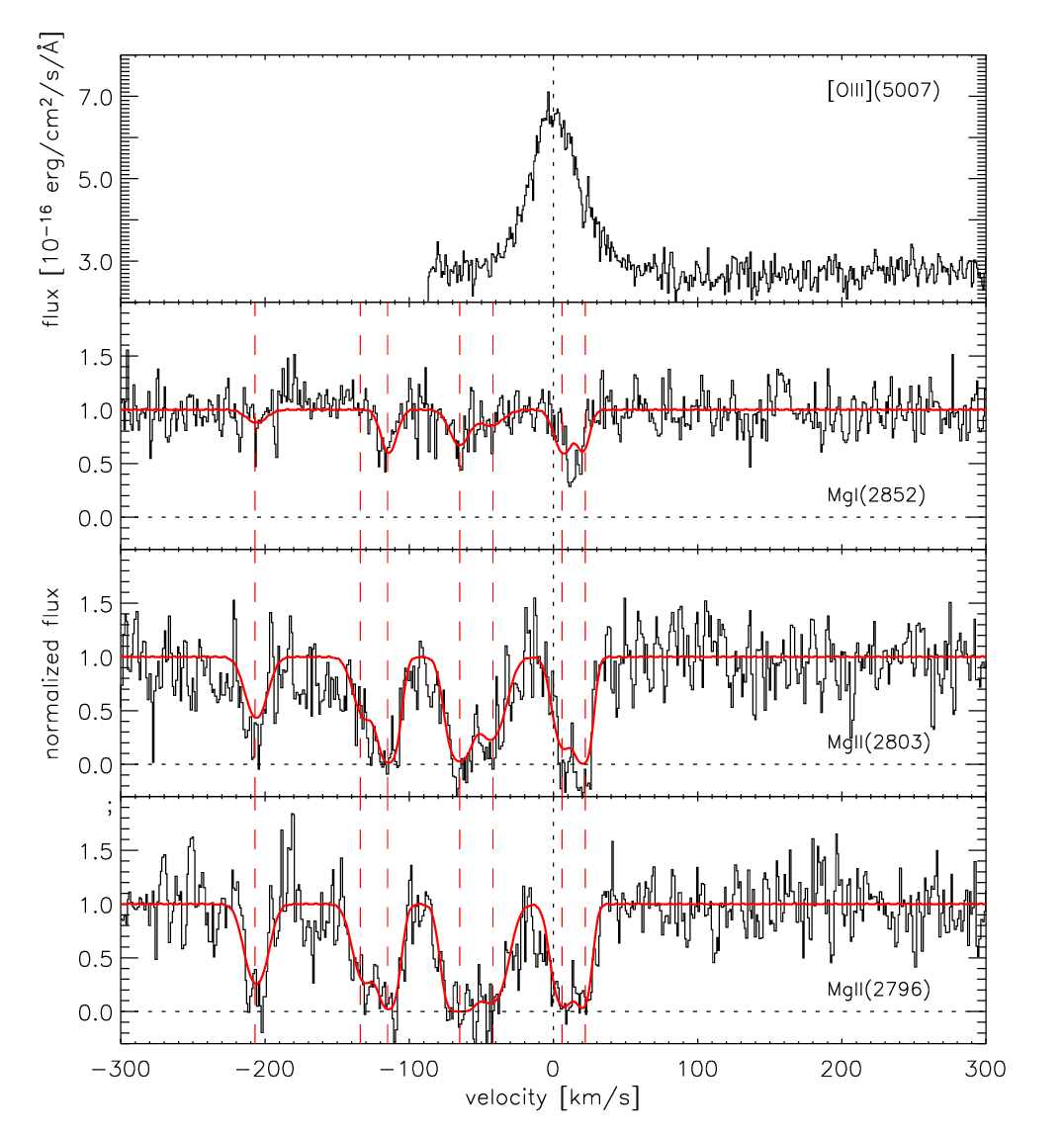}
\caption{Velocity distribution of the seven components in the Mg\,{\sc ii}-doublet and six in Mg\,{\sc i} and the corresponding fit to the different components with FIT/LYMAN. v=0 km/s corresponds to the redshift determined by the emission lines from the host.
We used only the first sample of the first night as this had the highest continuum and therefore the best statistics for the absorption lines. There was also no change visible in the fits from the other samples. \label{Magnesium}}
\end{figure*}

\begin{deluxetable*}{lccccccc}
\tablewidth{0pt} 
\tablecaption{Different components in the host
absorption lines} \tablehead{\colhead{Component} & \multicolumn{2} {c} {MgII[2796]} & \multicolumn{2} {c} {MgII[2803]} & \multicolumn{2} {c} {MgI[2852]}\\
 &  \colhead{b [km/s]} & \colhead{log N [cm$^{-2}$]} & \colhead{b [km/s]} & \colhead{log N [cm$^{-2}$]} &  \colhead{b [km/s]} & \colhead{log N [cm$^{-2}$]}& \colhead{N(MgII)/N(MgI)}}
 
\startdata
v$_1$ = +22 km/s & 4.7$\pm$0.8 & 12.91$\pm$0.26 & 4.7$\pm$0.8 & 13.50$\pm$0.24 & 4.7$\pm$0.8 & 11.53$\pm$0.13 & 2.07\\
v$_2$ = +6 km/s & 7.7$\pm$0.9 & 13.01$\pm$ 0.11 & 7.7$\pm$0.9 & 13.10$\pm$0.07 & 7.7$\pm$0.9 & 11.75$\pm$0.09 & 1.61\\
v$_3$ = -42 km/s & 11.1$\pm$1.4 & 13.06$\pm$0.07 & 11.1$\pm$1.4 & 13.13$\pm$0.06 & 11.1$\pm$1.4 & 11.38$\pm$0.18 & 2.02\\
v$_4$ = -65 km/s & 8.3$\pm$1.2 & 13.53$\pm$0.29 & 8.3$\pm$1.2 & 13.41$\pm$0.10 & 8.3$\pm$1.2 & 11.66$\pm$0.10 & 2.12\\
v$_5$ = -115 km/s & 6.4$\pm$0.9 & 13.06$\pm$0.18 & 6.4$\pm$0.9 & 13.42$\pm$0.20 & 6.4$\pm$0.9 & 11.68$\pm$0.10 & 1.90\\
v$_6$ = -134 km/s & 9.1$\pm$1.3 & 12.70$\pm$0.06 & 9.1$\pm$1.3 & 12.81$\pm$0.08 & \nodata & $<$6.5 & $>$6.6\\
v$_7$ = -207 km/s & 9.0 & 12.71$\pm$0.05 & 9.0 & 12.80 & 9.0 & 11.20$\pm$0.23 & 1.86
\enddata

\label{components}
\tablecomments{Values with no error bars indicate that they had to be fixed in order to achieve a reasonable fit.}
\end{deluxetable*}

For the absorption lines in the host galaxy, we found seven different velocity components for the two lines of the Mg\,{\sc ii} doublet, whereas Mg\,{\sc i} only shows six because of its lower line strength. The components span a velocity range of about 260 km/s (see Fig. \ref{Magnesium}) in both Mg\,{\sc ii}  and Mg\,{\sc i}. Mg\,{\sc i} shows nearly the same velocity differences as the Mg\,{\sc ii} components, so it seems that both arise from the same structure. This does not necessarily have to be the case due to the different ionization stages. The fit for the first two clouds however, does not match very well the Mg\,{\sc i} components, which might point to a different spatial origin for this component. The b parameter of the seven components is in the range of 5$-$11 km/s which is relatively narrow compared to QSO-Mg absorbers with similar EWs \citep{Churchill03}.\\
No time variability of the absorption lines was seen between the different epochs and we therefore only used the first epoch for the fitting due to the better S/N for the absorption lines. This rules out an origin in the vicinity of the burst \citep{PernaLoeb, Prochaska06}, where one would expect a considerable weakening of the Mg absorption between the first and the fourth night. Furthermore, the Mg\,{\sc ii} absorption in 5 of the 7 components goes down to zero, supporting the suggestion that the absorption does not occur at a very short distance from the central source. Otherwise, scattered light could lead to a covering factor smaller than one \citep{Savaglio04}.\\
The individual components at the different velocity shifts have similar column densities, though comparison of the ratios N(Mg\,{\sc ii}) to N(Mg\,{\sc i}) show that the conditions in the different absorbing structures might vary slightly. The total Mg\,{\sc ii} column density is 10$^{14.1\pm0.1}$cm$^{-2}$ which is in the regime of DLAs \citep{RaoTurnshek}. The gas in DLAs is predominantly neutral which is supported for our GRB by the presence of a relatively large column density of Mg\,{\sc i} (10$^{12.3\pm0.1}$ cm$^{-2}$) compared to N(Mg\,{\sc ii}). This ratio is also higher than in QSO-Mg absorbers \citep{Churchill03}.\\

\subsection{Interpretation of the Mg\,{\sc i} \& {\sc ii} velocity structure}
Mg\,{\sc i} and {\sc ii} have been detected in many GRB afterglow spectra \citep{Castro00, Fiore05, Starling05, Penprase06}, where two types of velocity spreads were found. The first type shows saturated Mg lines with a velocity broadening up to 200 km/s, comparable to the instrumental resolution in low resolution spectra or resolved in high resolution spectra as a truly broad absorption feature as in \cite{Penprase06}. The broad features from low resolution spectra could sometimes be partially resolved as in \cite{Castro00}, who found a second component with a $\Delta$v of 180 km/s. Therefore, it is likely that the broad lines sometimes actually consist of several distinct components, which could have been resolved with higher resolution spectrographs. The other type shows many high velocity components with $\Delta$v up to 3000 km/s as in \cite{Fiore05} and \cite{Starling05}.\\
The high velocity components are usually explained as being connected to the GRB itself like tracing nebula shells from former mass losses of the progenitor star, excited by the GRB \citep{Mirabal02} or strong winds which blow material of the star away from the Wolf-Rayet progenitor. The latter scenario was suggested for the high velocity components in GRB\,021004 \citep{Starling05, Fiore05, Chen05}. They found both broad and narrow structures in all the strongest absorption lines in UVES spectra of GRB\,021004 and GRB\,020813 up to velocities of 3000\,km/s. The velocities measured in GRB\,030329, however, are much lower, such winds can therefore not be the explanation for the components seen in our spectra. If those high velocity features were common for GRBs, they should have also been detected in GRB\,030329, in addition to the low velocity components, but there are no absorption features visible at comparable velocity distance from the main component. Neither were they detected in any other spectra than those of GRB\,021004, GRB\,020813 and recently GRB\,050505 \citep{Berger05} and GRB\,051111 \citep{Penprase06}, though the distances are far enough to be resolved in low resolution spectra as in the case of GRB021004 \citep{Mirabal02}. Either the ISM around the burst must be quite dense to produce these clumps, not all GRB progenitors have high mass losses and strong winds or the reason for these components is something completely different.\\
In some recent bursts, fine-structure lines of Si\,{\sc ii}, Fe\,{\sc ii}, C\,{\sc ii} and C\,{\sc iv} could be detected as first reported by \cite{Vreeswijk04} in a GRB spectrum. Fine-structure lines have only been found in high density environments such as $\eta$ Carinae \citep{Gull05}. It is assumed that in GRBs, these lines come from electron excitation in the hot, turbulent vicinity of the burst. They are however not the reason for the structures in the Mg lines in our spectrum as these transitions are extremely weak for Mg \citep{Morton03}.\\
Another possibility for the origin of the Mg lines in the spectrum of the GRB\,030329 afterglow are supergalactic winds from starburst regions as found in spectra of low-redshift starburst galaxies and clearly visible e.g. in H$\alpha$ light from M82. These winds were discovered in X-ray data, as the heated ISM emits X-rays. \cite{Heckman00} studied the structures of NaD lines from such starburst galaxies. They found blueshifted components relative to the host galaxy up to 600 km/s with Doppler width of more than 100 km/s which seems likely for a hot environment (T$\approx$70,000 K). This however rules out the possibility of an active starburst wind for our absorption line features as the velocities are too low for a wind origin and in addition, the lines are very narrow, so the absorbing atoms cannot be in a hot environment.\\
These winds might however not be the end of the story. In the MW as well as in other galaxies, so-called ``high velocity clouds«« have been found, consisting mainly of HI, but O\,{\sc vi} has also been discovered from the same region \citep{Fox05}. This gave rise to the theory of the galactic fountain where starburst regions blow hot, ionized material out into the halo which then cools down and eventually falls back again onto the galaxy. The Mg absorption lines in our spectra might therefore be caused by some intermediate stages of that process, when the starburst wind has cooled and slowed down. The fact that, except for the two lowest-velocity systems, the velocities are larger than the rotational velocity of the host galaxy as determined from the emission line width, indicates that they are most likely not just related to large structures in the host galaxy like spiral arms. With the escape velocity typically being less than 50\%  larger than the mean rotational velocity, half of our velocity systems are likely to be unbound to the gravitational potential of the host galaxy. The star burst wind suggestion is therefore an appealing scenario for those structures.\\
The same process might cause Mg\,{\sc ii} absorbers, which have often been found in the sightline of  QSOs \citep{Ellison} with large EW.  Similar structures have also been discovered in Lyman break galaxies (LBG) \citep{Adelberger03} but only for C\,{\sc iv} absorption lines. DLAs associated with Mg-absorbers show black bottom saturated Mg velocity components with narrow spreads of about 40 to 60 km/s. So-called ``doubles«« have smaller EWs and highly complex features separated by a large $\Delta$v \citep{Churchill99} with their individual components having Doppler widths comparable to those found in GRB 030329. This might lead to the conclusion that Mg QSO absorbers and GRB absorbers have the same origin but that they are probing different regions of the galaxy. \cite{Petitjean05} also suggested that QSO-Mg absorbers could arise in optically thick clouds in the galactic halos, created by material ejection from starburst regions, being the remnants of a former starburst wind.\\
For strong QSO absorbers, there usually exist velocities red- and blueshifted compared to the central redshift of the intervening galaxy \citep{Bond} whereas in our case mainly blueshifted lines exist, except for the first two components. This might be a hint to the position of the burst in the galaxy. If these components are really due to some metal rich clouds in or around the host galaxy, the burst could lie slightly behind the center plane of the galaxy or the star burst cloud which causes the few redshifted components, whereas the rest of the lines come from material flowing away from the burst site. \\

\section{Conclusion}
In this paper, we have presented high resolution spectra of a typical long-duration GRB at z=0.16867, which had the rare case to contain both emission lines from the host galaxy as well as absorption lines from the surrounding medium of the burst. The host galaxy is a subluminous (m$_\mathrm{V}$=$-$16.37), low mass (M$_\star$ = 10$^{7.75\pm0.15}$M$_\odot$) starburst with a high SFR/L of 14.1 M$_{\odot}$ yr$^{-1}$ L$_*^{-1}$ or log SFR/M = $-$8.5 yr$^{-1}$ compared to other galaxies at this redhsift. This confirmes the values of the SSFR found by \cite{Hjorth03} and \cite{Sollerman05} from late FORS spectra of the host galaxy. Furthermore, the high resolution UVES spectra made it possible to derive a strong upper limit for [N II] and therefore to place the value of the metallicity in the lower branch with log(O/H)=7.8 or 1/7 Z$_\odot$.\\
In absorption, only Mg\,{\sc ii} and Mg\,{\sc i} from the host could be detected. Due to the high resolution again, however, we were the first to resolve these usually broad absorption lines into six and seven narrow velocity components for Mg\,{\sc i} and Mg\,{\sc ii} respectively. These components span a relatively small range of velocities up to v$\sim$260 km/s, which can usually  not be resolved in low resolution spectra. The majority of these components are however outside of the galaxy whose emission lines have only a width of 55 km/s. Both Mg\,{\sc i} and Mg\,{\sc ii} seem to originate in the same clouds which are predominantly neutral with N(Mg\,{\sc ii})/N(Mg\,{\sc i}) of around 2. Structures similar in width and velocity distance have been found in QSO Mg absorbers. They could be clouds in the halo of the host galaxy, possibly from material ejected from the inner regions of the galaxy by former starburst driven superwinds which have already slowed down.\\
The {\it Swift} satellite now provides faster information of the burst position and more early, high resolution spectra can be obtained. Very early spectra, compared with later times, could also reveal possible changes of the absorption lines due to ionization of the ISM by the burst \citep{Vreeswijk06b}.

\acknowledgements

C.T. wants to thank Johan Fynbo for reading the manuscript and discussions about upper limits as well as Alexander Kann for editing the text and the references. Thanks also to Klaas Wiersema for the preprint of his paper on 060218 and informations about advanced line diagnostics. The ``DARK Cosmology Centre«« is funded by the Danish National Research Foundation

\end{document}